# Neutron Repulsion

**O. Manuel***

*Associate, Climate & Solar Science Institute*
*833 Broadway, 104, Cape Girardeau, MO 63701*
*Emeritus Professor, Nuclear and Space Studies*
*University of Missouri, Rolla, MO 65401*
*Former NASA Principal Investigator*
*For the Apollo Mission to the Moon*
*Websites: http://www.omatumr.com*
*http://myprofile.cos.com/manuelo09*

Earth is connected gravitationally, magnetically and electrically to its heat source - a neutron star that is obscured from view by waste products in the photosphere. Neutron repulsion is like the hot filament in an incandescent light bulb. Excited neutrons are emitted from the solar core and decay into hydrogen that glows in the photosphere like a frosted light bulb. Neutron repulsion was recognized in nuclear rest mass data in 2000 as the overlooked source of energy, the keystone of an arch that locked together these puzzling space-age observations: **1.)** Excess $^{136}$Xe accompanied primordial helium in the stellar debris that formed the solar system (Fig. 1); **2.)** The Sun formed on the supernova core (Fig. 2); **3.)** Waste products from the core pass through an iron-rich mantle, selectively carrying lighter elements and lighter isotopes of each element into the photosphere (Figs. 3-4); and **4.)** Neutron repulsion powers the Sun and sustains life (Figs. 5-7). Together these findings offer a framework for understanding how: **a.)** The Sun generates and releases neutrinos, energy and solar-wind hydrogen and helium; **b.)** An inhabitable planet formed and life evolved around an ordinary-looking star; **c.)** Continuous climate change - induced by cyclic changes in gravitational interactions of the Sun's energetic core with planets - has favored survival by adaptation.

*E-mail: omatumr@yahoo.com or omatumr2@gmail.com

# Introduction

As an expression of gratitude for 50 joyful years of "truthing" since starting research in 1960, I was pleased to accept the invitation to review evidence of neutron repulsion and its implications for the evolution of life. It will be shown that life [1] and atomic nuclei have evolved together on opposite sides of the Sun's opaque photosphere.

My conclusions do not support Fowler's concerns that solar neutrinos and the chemical composition of living organisms violate basic concepts of nuclear astrophysics: *"Indeed there are details to be attended to, but they are overshadowed by serious difficulties in the most basic concepts of nuclear astrophysics. On square one, the solar neutrino puzzle is still with us . . . indicating that we do not even understand how our own star really works. On square two we still cannot show in the laboratory and in theoretical calculations why the ratio of oxygen to carbon in the sun and similar stars is close to two-to-one . . . We humans are mostly (90%) oxygen and carbon. We understand in a general way the chemistry and biology involved, but we certainly do not understand the nuclear astrophysics which produced the oxygen and carbon in our bodies."* [2].

Fusion of three or four nuclei of helium ($^{4}$He) produced the most abundant isotopes of carbon ($^{12}$C) and oxygen ($^{16}$O), respectively, as Fowler assumed [2], but material in the photosphere will be shown to be like smoke pouring from the solar chimney: Waste products from the nuclear reactions that power the Sun and sustain life on Earth. Abundances of oxygen, carbon and other elements in the photosphere are severely mass fractionated. Major by-products of solar luminosity, hydrogen and helium, comprise respectively ~91% and ~9% of all atoms there, but abundances of oxygen and carbon exceed those of hydrogen inside the Sun. The ratio of these two He-fusion products in the interior of the Sun does not violate predictions of nuclear astrophysics.

Neutron repulsion is the triumphant arch through which the puzzling oxygen to carbon ratio in the photosphere and many other baffling observations about the solar system and its heat source - the Sun - could be viewed as pieces of a surprisingly simple mosaic on the origin, chemical composition, and source of nuclear energy that gave birth to the solar system and has continued to power the Sun and sustain life on planet Earth.

First it will be helpful to review the crucial - and totally unexpected - 1975 experimental observation that changed opinions about the origin of the solar system. Figure 1 shows isotope (vertical) and elemental (horizontal) abundance data [3,4] from analysis of noble gases in mineral separates of the Allende meteorite. The link of primordial helium with excess $^{136}$Xe from the stellar r-process [5] suggests that elements were made locally [6], rather than in remote stars that injected them into the interstellar medium from which an interstellar cloud eventually collapsed to form the solar system. Local element synthesis also explains why decay products of short-lived isotopes [7-14] and isotopic anomalies from nucleosynthesis were seen [15,16] in the Earth and in diverse meteorites. The Allende meteorite sampled two primordial reservoirs of xenon (**Xe-1** and **Xe-2**) from the inner and outer layers, respectively, of a local star [6,17]. ***All*** primordial helium was initially in the reservoir with "strange" **Xe-2**; little or none was with "normal" **Xe-1**.

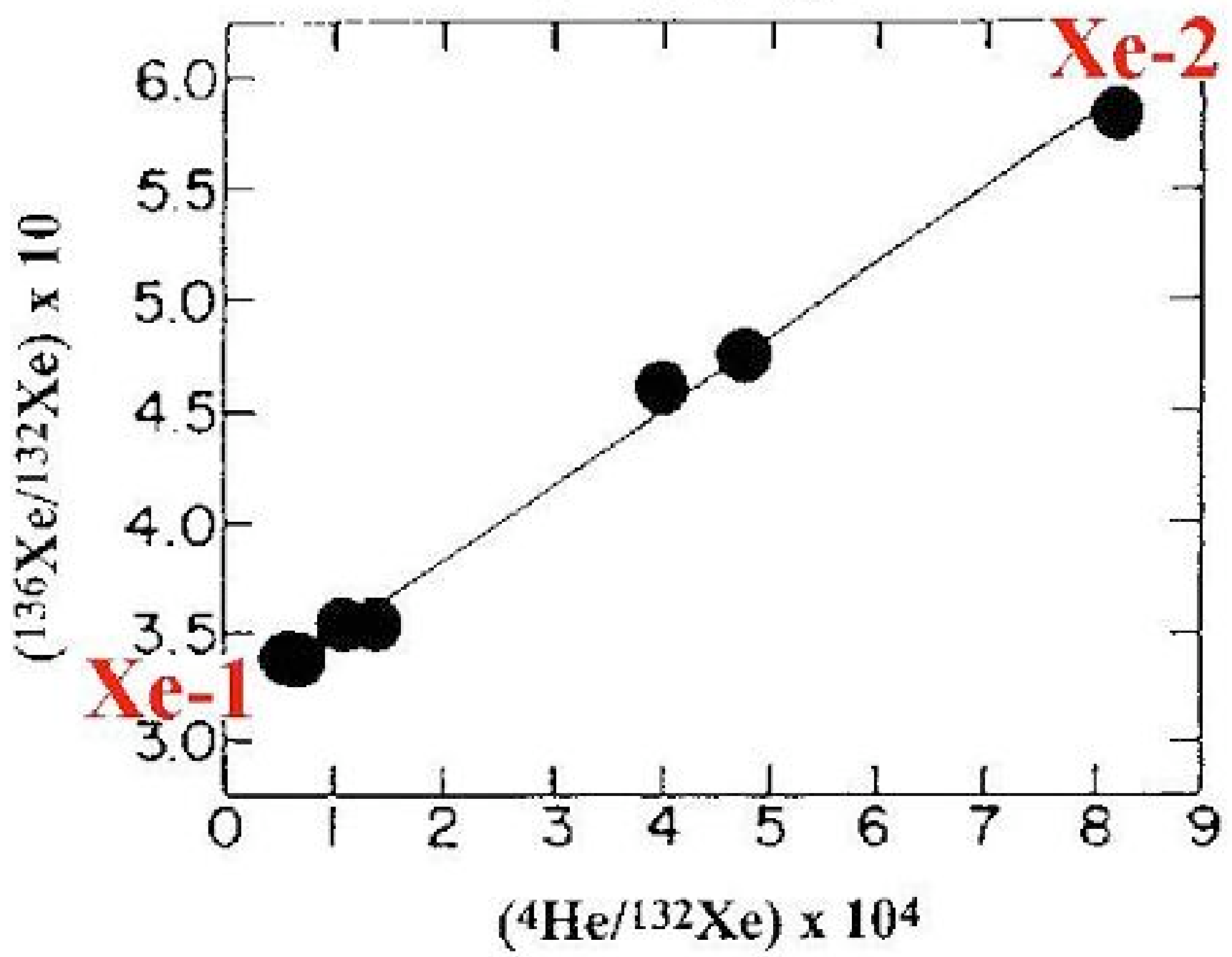

***Fig. 1***. *The association of all primordial helium with excess $^{136}$Xe in "strange" xenon (**Xe-2**), and its absence from the "normal" xenon (**Xe-1**) component, was the first clear indication that elements and isotopes never completely mixed in the stellar debris that formed the solar system [6,17]. **Xe-1** and **Xe-2** came respectively from the inner and outer layers of the local supernova that gave birth to the solar system.*

Figure 2 shows the scheme proposed for birth of the solar system [6,17-19] to explain the noble gas data in Figure 1 [3,4] and other puzzling observations [7-16]. It assumes that the elements came from a star that had evolved in the manner described by B2FH [5].

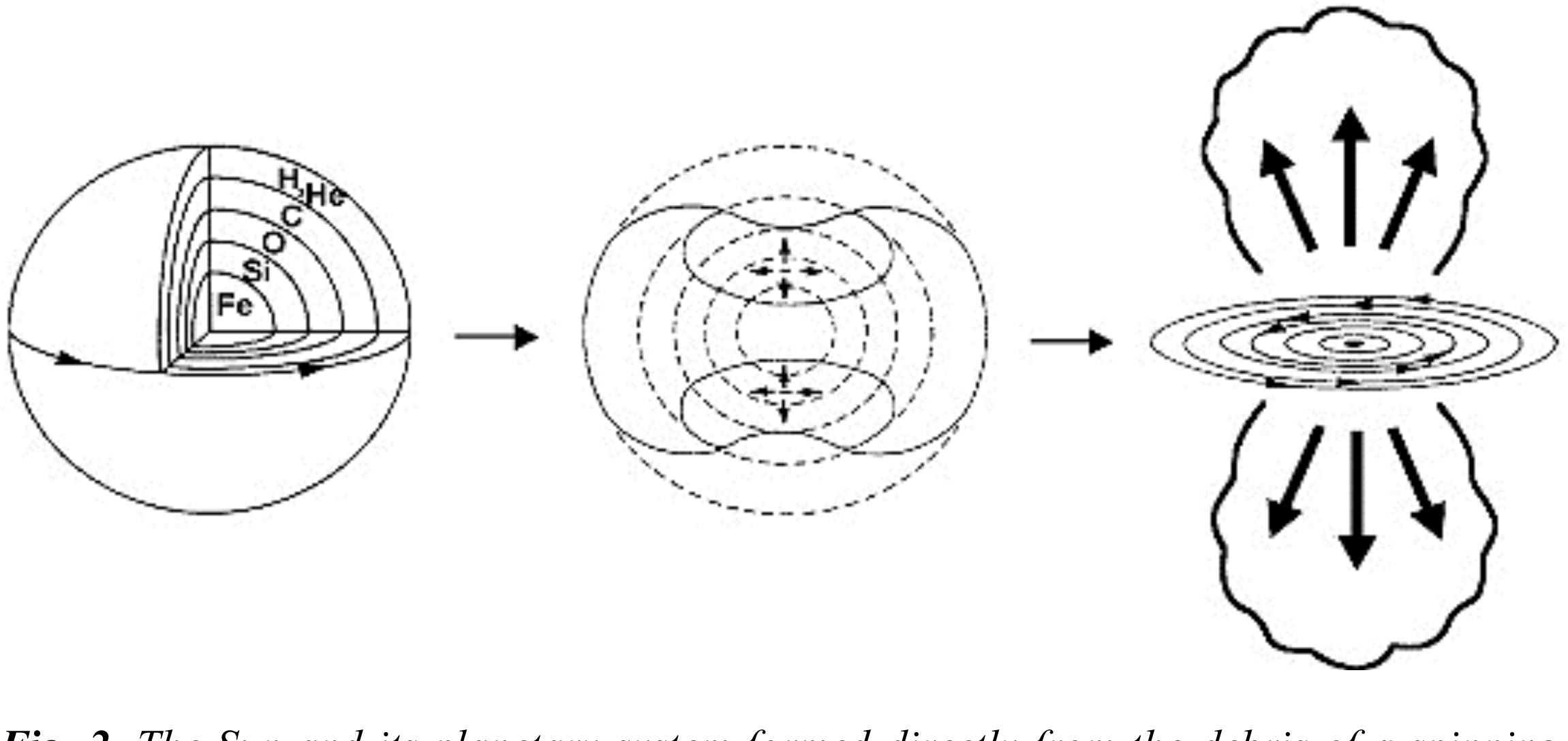

***Fig. 2.*** *The Sun and its planetary system formed directly from the debris of a spinning, evolved [5] star that imploded and then exploded axially to produce a planetary disk in the equatorial plane orbiting the remnant neutron-rich core on which the Sun formed.*

The Figure 2 scenario for the birth of the solar system was presented at the 1976 AGU [17] and ACS [18] national meetings and at the 1977 Robert A Welch Conference on Cosmochemistry [19], and debated there [18-21] and in *Science* [22,23]. Advocates of *in situ* super-heavy element fission as the source of excess $^{136}$Xe in meteorites published a supporting study in the *Proceedings of the National Academy of Science* [24]. Another group proposed that excess $^{136}$Xe might have been injected from a nearby supernova [25], without addressing the ubiquitous link of primordial helium with excess $^{136}$Xe (Figure 1).

Many studies [26-56] over the next few years confirmed that meteorites and planets trapped fresh, poorly mixed nucleosynthesis products as they formed. Isotope analysis revealed levels of $^{16}$O that distinguished six, chemically distinct types of meteorites and planets: *"Objects of one category cannot be derived by fractionation or differentiation from the source materials of any other category"* [26]. Nucleogenetic isotopic anomalies were observed in diverse elements [26-28,30-37,39-42,44-46,48-56]. The decay product of yet another short-lived isotope, $^{107}$Pd, was discovered in meteorites [38].

New evidence was found of linked chemical and isotopic anomalies in the material that formed the solar system [26,42,44,56]. Another study confirmed [43] that iron meteorites formed as early as "primitive" meteorites [11]. The Earth's inventory of radiogenic and primitive noble gases was shown to be consistent with Earth's accretion in layers beginning with its iron core [52], as Turekian and Clarke [57] and Vinogradov [58] had suggested earlier. Isotopic anomalies were also identified from a local irradiation [29,41,54,55], as first suggested by Fowler, Greenstein and Hoyle [59].

Professor Begemann [45] summarized the situation in 1980 by stating that recent measurements do not agree with the *"classical picture of the pre-solar nebula"* as *"a hot, well-mixed cloud of chemically and isotopically uniform composition."*

As measurement after measurement [26-56] accumulated on one side of the debate [17-23] over the formation of the solar system (Fig. 1), it seemed that the matter might be settled in 1983 if two seemingly intractable puzzles could be solved: a.) The Sun seemed to be a giant "ball of hydrogen" in the middle of iron-rich supernova debris. b.) Severely mass fractionated isotopes had been reported in meteorites and in lunar samples since 1960 [7,8,15,27,28,47,60-70], but the fractionation site had not been identified. A single solution for both puzzles was published in a peer-reviewed paper in September of 1983.

Earlier in March at the 14th Lunar & Planetary Science Conference, proponents of *in situ* fission for excess $^{136}$Xe (also called CCFXe) started to recant [71,72]. E.g., *". . . the present data seem to rule out the in situ fission model*" in one [71] of two papers coauthored with researchers from other universities, and *". . . the apparent association with nitrogen of* $\delta^{15}N = -273‰$ *argues strongly for a nucleogenetic origin of CCFXe, . . ."* in the other [72]. At the same conference we reported that isotope abundances in the solar wind showed evidence of mass fractionation and formation of the solar system from *". . . chemically and isotopically heterogeneous"* supernova (SN) debris that *". . . contained short-1ived radioactivities"*, with the Sun's radiant energy *". . . generated, not by fusion, but by radiation from a hot SN core in the Sun's interior"* [73].

A 26 June 1983 news report in Nature on the 14th Lunar & Planetary Science Conference (LPSC), *"The demise of established dogmas on the formation of the Solar System"* [74], overlooked our LPSC paper [73] but acknowledged that the results presented in two other LPSC papers [71,72] *". . . have led the principal defendants in the argument over the origin of CCF-Xe to concur in favor of the supernova hypothesis."*

Our paper on solar mass fractionation and the Sun's iron-rich interior [75], based on the noble gas isotope data presented in Table 1 of the 14th LPSC paper [73], had been submitted for review. The manuscript was published as a peer-reviewed paper [75] in September of 1983. Figure 3 shows the empirical evidence for solar mass fractionation [73,75] that was identified by comparing the isotope ratios of noble gases in the solar wind [76] with those in planetary material [77-79].

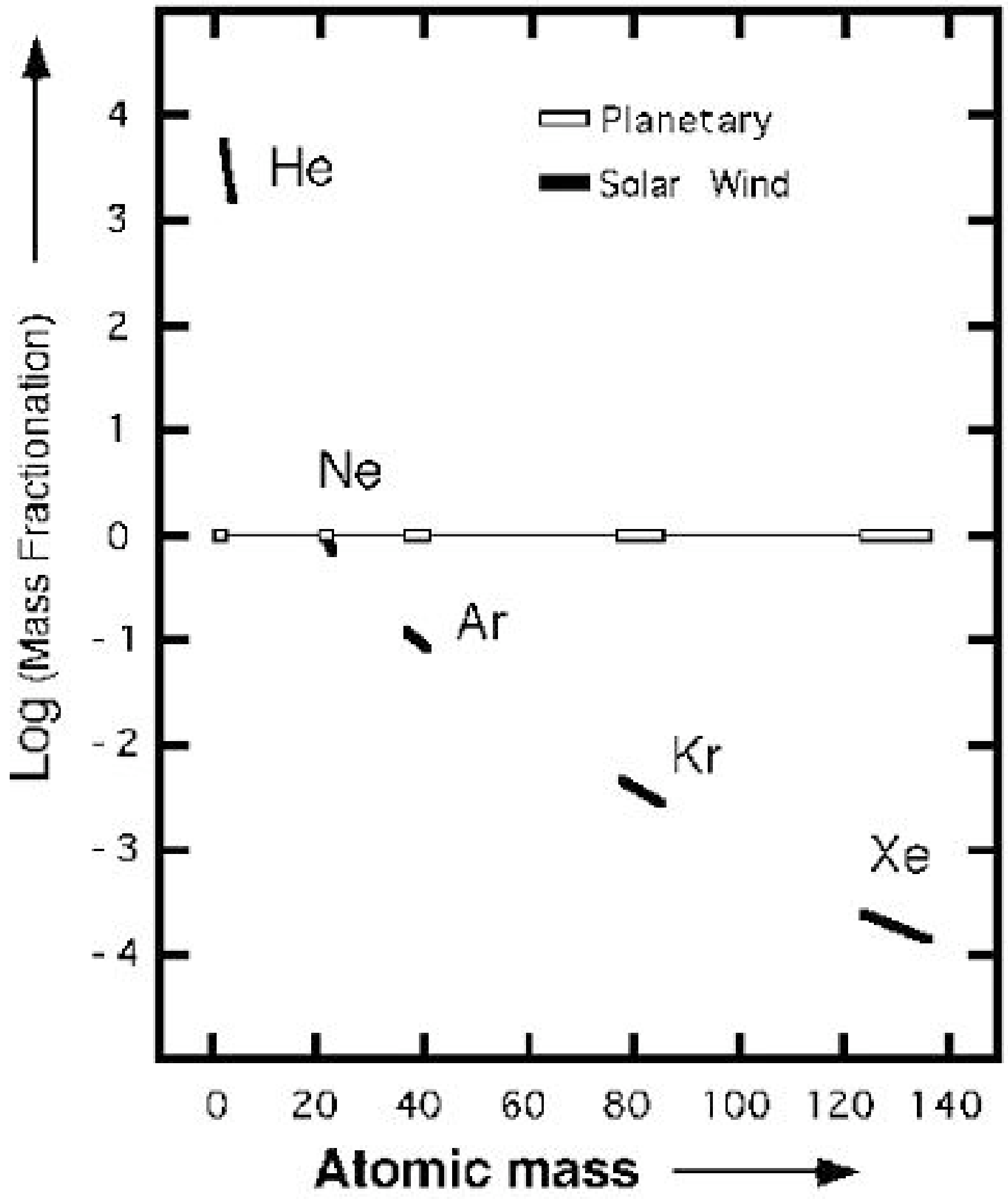

***Fig. 3.*** *Lighter isotopes of mass ($m_L$) are systematically enriched relative to heavier isotopes of mass ($m_H$) in noble gases coming from the top of the Sun by a factor, **f**, where empirically,* $f = [(m_H)/(m_L)]^{4.56}$ *[75].*

Figure 4 shows the composition that Manuel and Hwaung [75] calculated for material inside the Sun, after correcting abundances in the solar photosphere [80] for the ~9-stages of mass-dependent fractionation that had been observed across the isotopes of noble gases that were implanted in lunar soils from the Sun.

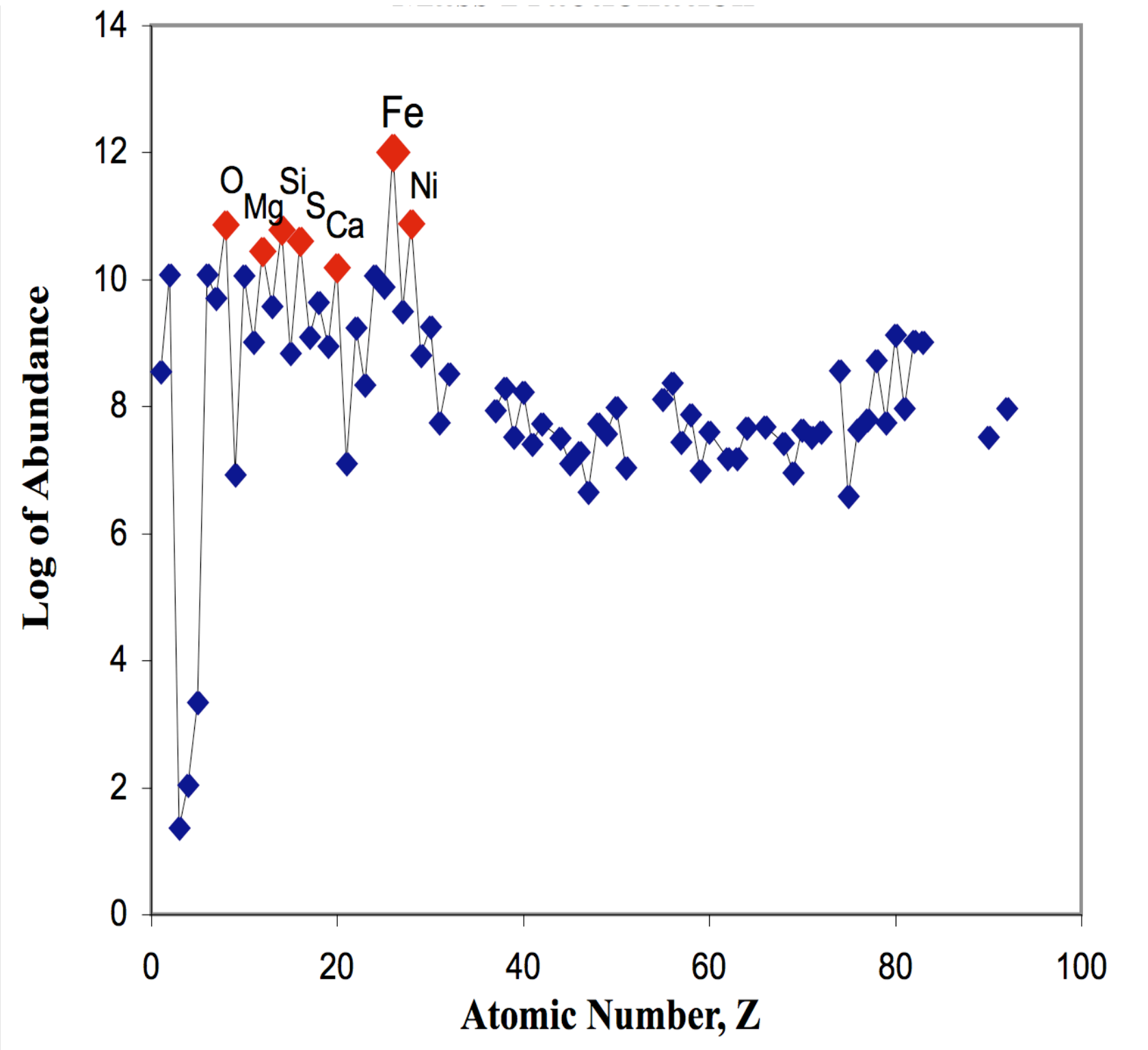

***Fig. 4.*** *Material in the Sun is mostly the same elements that constitute the matrix of rocky planets and ordinary meteorites [81]: Fe, O, Ni, Si, S, Mg and Ca represented with large diamonds. The probability of this agreement being a coincidence is essentially zero.*

Finally in December of 1983, the leading proponent of the super-heavy element fission hypothesis [3,4] coauthored a report [82] with *evidence against* the suggestion that decay of a super-heavy element [4] produced the excess $^{136}Xe$ (Fig. 1) in meteorites.

Still the scientific community did not take the iron Sun hypothesis [75] seriously until its source of luminosity was identified - despite an earlier report that the Sun might be a pulsar [83], many additional reports of nucleogenetic anomalies in meteorites and planets [e.g., 84-94], and indications of an iron-rich solar interior from helioseismology [95], the Galileo probe of Jupiter [96], and Wind spacecraft measurements on a solar flare [97].

Neutron repulsion was first recognized in nuclear rest mass data in 2000 and reported [98,99] as a source of solar energy that explained the "solar neutrino puzzle" [100], the separation of d- and l- amino acids [101], the supernova birth of the solar system (Fig. 2), and values reported for solar neutrinos, solar luminosity, solar wind hydrogen, and solar mass fractionation. Neutron repulsion will be explained in the next section as the source of energy that powers the Sun and sustains the evolution of life on Earth.

# Neutron Repulsion

Neutron repulsion is recorded as rest mass (potential energy) in every nucleus with two or more neutrons. Sixty-one years after the discovery of neutron-induced fission [102], neutron repulsion was recognized when five students who were enrolled in an Advanced Nuclear Chemistry class (Chem 471) - Cynthia Bolon, Shelonda Finch, Daniel Ragland, Matthew Seelke and Bing Zhang - helped the author develop three-dimensional (3-D) plots of reduced nuclear variables: Z/A (charge per nucleon) = [Atomic Number]/[Mass Number]; M/A (mass or potential per nucleon) = [Mass]/[Mass Number]) from the latest available nuclear rest mass data [103]. The results are shown as the "Chart of the Nuclides" on the left side of Figure 5.

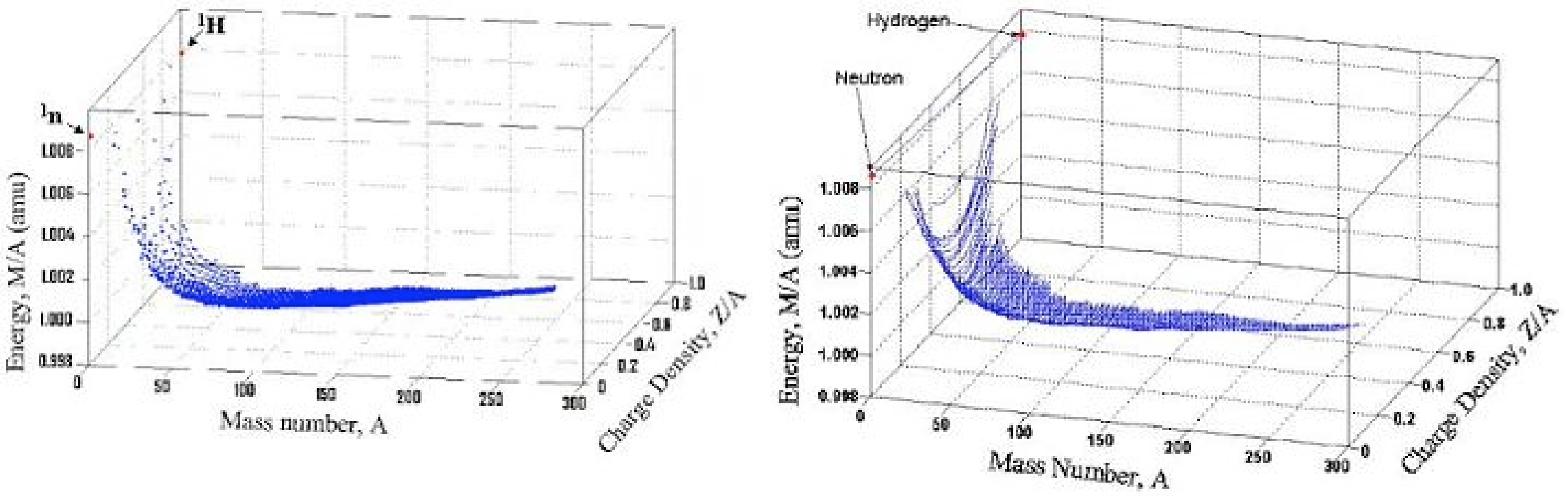

***Fig 5.*** *The author and five students developed the "Cradle of the Nuclides" on the left in the spring semester of 2000 [M = Mass in atomic mass units (amu); Z = Atomic number; A = Mass number]. The vertical axis is mass (potential energy) per nucleon (M/A), the horizontal axis is mass number (A), and the depth axis is charge density (Z/A) or charge per nucleon. Mass parabolas were fitted to the data points at each mass number (A) to produce the graph on the right. When these mass parabolas are extrapolated to the front panel, at Z/A = 0, neutron repulsion is revealed as rest mass. At Z/A ~ 0.5, attraction of neutrons for protons decreases the rest mass to produce ordinary nuclei [98].*

Prior to the discovery of neutron repulsion [98,99], neutron stars were considered "dead" nuclear matter, bound by ~93 MeV/neutron [104]. Recently large research consortia report that neutron stars are powerful sources of energy [105,106] and news reports that, *"the source of the pulsar's power may be hidden deep within its surface"* [107].

The potential energy (mass) per nucleon from repulsive interactions between neutrons can be seen more clearly in Figure 6 as intercepts of empirically defined mass parabolas with the front panel at Z/A = 0. Intercepts with the front panel show what the values of M/A would be if the nucleus were composed entirely of neutrons. Likewise intercepts of the mass parabolas with the back panel at Z/A = 1 show even higher values of M/A if the nucleus were composed entirely of protons. Coulomb repulsion between positive charges on protons adds to the potential energy (mass) per nucleon at Z/A =1. Differences between the values of intercepts with the front and back panels in Figure 6 arise entirely from Coulomb repulsion between positive charges on protons [108, 109].

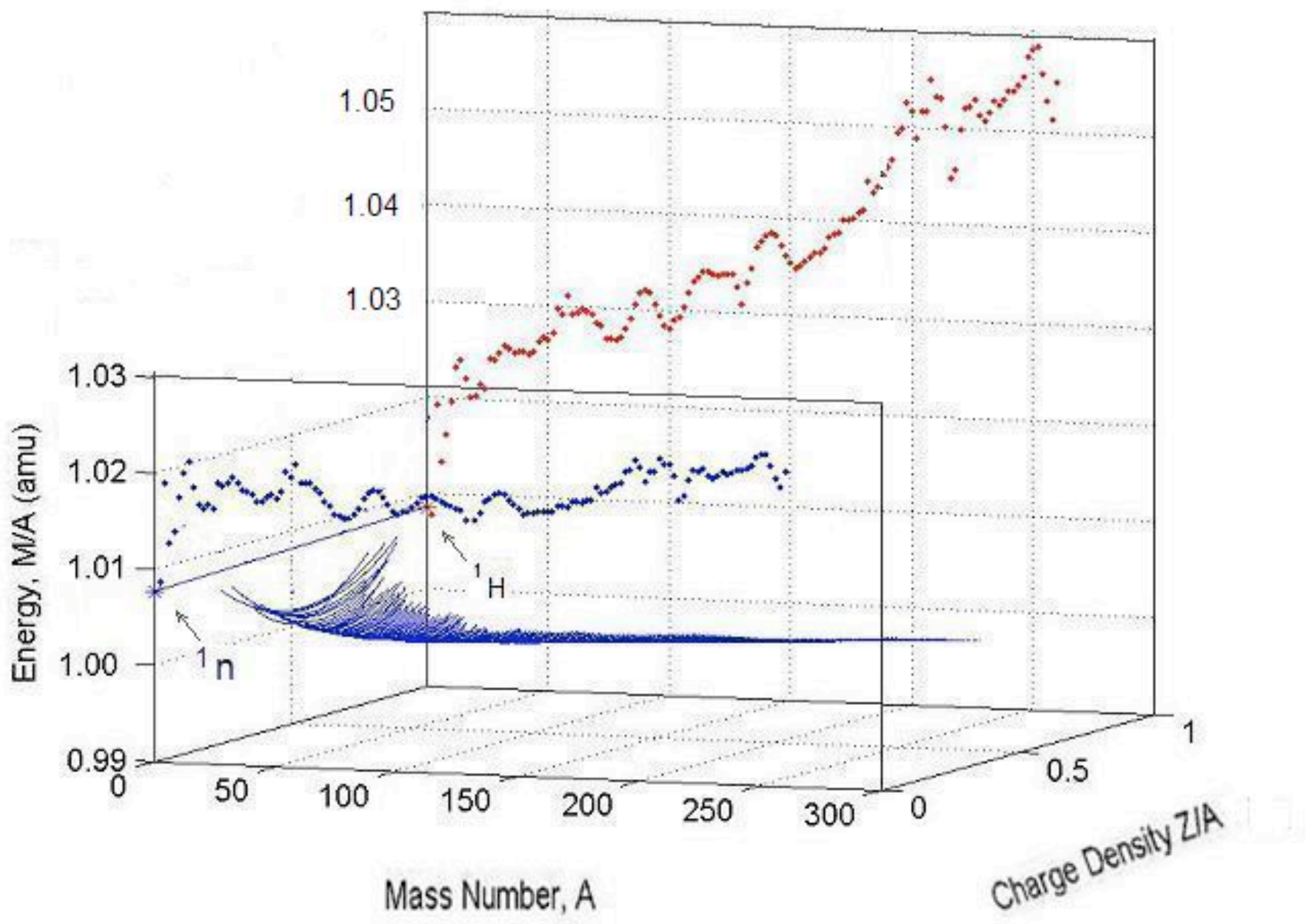

***Fig 6.*** *The potential energy per nucleon (M/A) from repulsive interactions between neutrons is shown as intercepts with the front panel at Z/A = 0 for mass parabolas fitted to nuclear rest mass data of ground state nuclides [103]. Coulomb repulsion between positive charges on protons explains quantitatively why values of intercepts with the back panel at Z/A = 1 become increasingly higher as the mass number, A, increases [108].*

Other researchers [110] independently concluded that useful information on neutron stars could be obtained by extrapolating atomic mass data out to *"homogeneous or infinite nuclear matter (INM)"* (ref. 110, page 1042).

It may be helpful to display the information in Figures 5 and 6 on a conventional 2-D graph that compares the potential energy per nucleon of ordinary nuclei – values of M/A for Z/A ~ 0.5 - with values of M/A calculated for homogeneous, infinite nuclear matter (INM) at the intercepts where Z/A = 0 and Z/A = 1. These results are shown in Figure 7.

In lower right side of Figure 7 it can be seen that fission of heavy nuclei involves small changes in nuclear stability and typically release ~ 0.1% of the rest mass as energy. Fusion of hydrogen into helium releases ~ 0.7% of the rest mass as energy. Complete fusion of hydrogen into iron would release ~ 0.8% of the rest mass as energy. Neutron emission from a neutron star would release 1.1-2.4% of the rest mass as energy [108]. If neutron-emission were followed by neutron-decay and then by fusion of the neutron decay product (hydrogen), as happens in the Sun, a total of about ~2-3% of the initial rest mass might be converted into stellar energy [98,99].

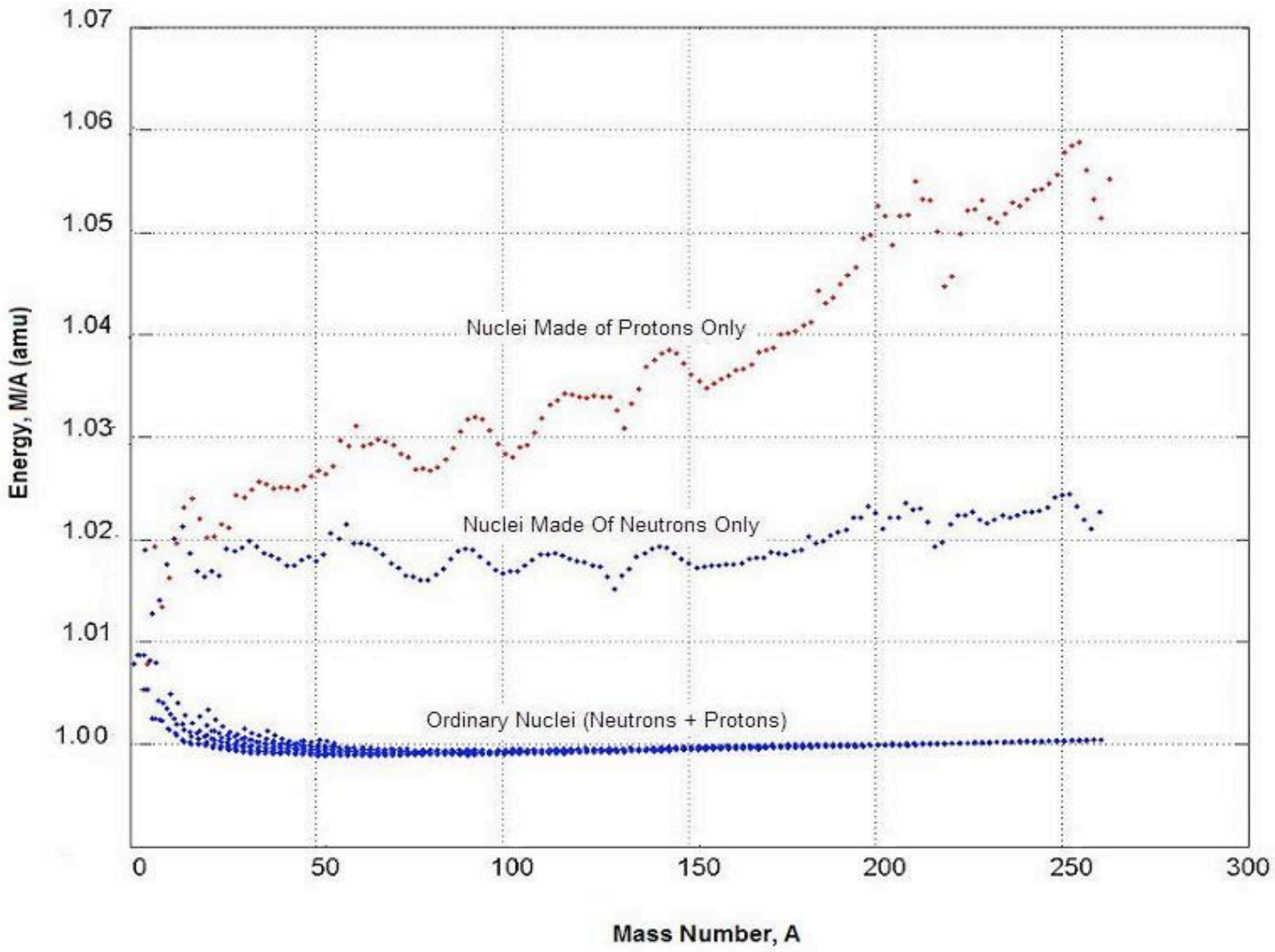

***Fig 7.*** *Most ordinary nuclei with Z/A ~ 0.5 lie along the lower part of this diagram and have values of M/A ~ 1.00 amu/nucleon. Light fusible nuclei have values of M/A ~ 1.00-1.01 amu per nucleon. Nuclei made of neutrons (Z/A ~ 0) only have M/A ~ 1.02. Nuclei made of protons only would have even greater potential energy per nucleon because of Coulomb repulsion [108].*

As can be seen in Figure 7, the material proposed [111, 112] over seventy years ago for neutron cores of stars consists of excited neutrons with M/A ~ 1.02-1.03 amu/nucleon and Z/A ~ 0. Neutron emission would release ~10-22 MeV of energy [108,109]. After emission, the free neutron would be expected to decay spontaneously to hydrogen in ~10 minutes, unless first captured by another nucleus.

Since material in the interior of the Sun [73,75] shows no compelling evidence of neutron-capture there, we proposed at the 32nd Lunar & Planetary Science Conference (LPSC) in March 2001 that solar luminosity and solar wind hydrogen might be explained, and the solar neutrino puzzle solved, by a series of nuclear reactions triggered by neutron repulsion and neutron emission from the core of the Sun [99]:

1. Neutron emission: $<{}_0^1n> \rightarrow {}_0^1n + \sim 10\text{-}22$ MeV
2. Neutron decay: ${}_0^1n \rightarrow {}_1^1H^+ + e^- + \overline{\nu_e} + 0.78$ MeV
3. Migration up and fusion of $H^+$: $4\ {}_1^1H^+ + 2\ e^- \rightarrow {}_2^4He^{++} + 2\ \nu_e + 26.73$ MeV
4. Excess $H^+$ released in solar wind: $3 \times 10^{43}\ H^+/yr \rightarrow$ Depart in solar wind

Our request to make an oral presentation at the 32nd LPSC was denied, but the poster was well attended and we published the details in other papers [e.g., 98, 108,109].

A few months after the 32nd LPSC, the Sudbury Neutrino Observatory research consortium reported new measurements that suggested oscillations as the solution to the solar neutrino puzzle [113]. The total flux of all flavors of solar neutrinos were reported to be *". . . in close agreement the predictions of solar models"* [Abstract, reference #113]. The next year the group reported direct evidence for solar neutrino flavor transformations from neutral-current interactions [114]. Again our papers were not cited, but the paper concluded even more forcibly that the new results are ". . . *in agreement with the SSM prediction*" [Last sentence of summary, reference # 114, SSM = Standard Solar Model].

The importance of these new measurements is not evidence of solar neutrino oscillations and/or the Standard Solar Model [SSM], but <u>confirmation that the flux of solar electron neutrinos is ~35% of that expected if H-fusion alone generates solar luminosity</u>. Doubt was cast on the original interpretation of the measurements by a later study that detected even more oscillations than would be possible if there were three flavors of neutrinos [115]. In the accompanying news report [116], author Professor Bryon Roe said, *"These results imply that there are either new particles or forces we had not previously imagined,"* and *"The simplest explanation involves adding new neutrino-like particles, or sterile neutrinos, which do not have the normal weak interactions."*

In view of improved limits on the flux of solar electron neutrinos reported by the Sudbury Neutrino Observatory [113,114], equations (1)-(4) can now be rewritten to show more clearly the percent (%) of solar energy (SE), solar neutrinos (SN) and solar wind hydrogen (SW) generated in transforming a neutron, from inside the central neutron star, to a helium atom in the solar wind:

| | | | |
|---|---|---|---|
| 1'. Neutron emission: | | $<{}_0^1n> \rightarrow {}_0^1n$ + ~12 MeV | **~60% SE** |
| 2'. Neutron decay: | | ${}_0^1n \rightarrow {}_1^1H^+ + e^- + \overline{\nu_e}$ + ~1 MeV | **~05% SE** |
| 3'. Hydrogen fusion: | **~100% SN** | ${}_1^1H^+ \rightarrow {}_2^4He + 0.5\ \nu_e$ + ~7 MeV | **~35% SE** |
| 4'. Discharge excess Hydrogen: | | $3 \times 10^{43}\ H^+/yr \rightarrow$ Departs in SW | **~100% SW** |

Neutron-emission, neutron-decay & H-fusion → **~100% SN + ~100% SE + ~100% SW**

Before concluding the discussion of neutron repulsion, it should be noted that the scheme in Figure 2 [17-19] - the axial explosion of the star that gave birth to the solar system – was an attempt to explain how a star that evolved according to B2FH [5] might explode and leave the heterogeneous stellar debris seen in the Allende meteorite [3] (Figure 1).

It now seems more likely that asymmetric stellar explosions, bipolar outflows [117] and flares [118] are powered by neutron repulsion [98] and guided by the magnetic field of the neutron core [112] or the iron-rich material that surrounds it [118]. Neutron decay into oppositely charged H+ and e-, in the presence of a strong magnetic field, may initiate the electrical currents that have long puzzled careful observers of the Sun and other stellar objects [106,119,120].

## Conclusions

Dynamic competition between gravitational attraction and neutron repulsion sustains our dynamic universe, the Sun, and life on planet Earth. Nuclear matter in the solar system is mostly dissociating rather than coalescing (fusing together). As shown in the above table, the potential energy per nucleon in the solar core is almost twice that available from hydrogen fusion. If the bulk of the Sun's mass is in a central neutron star and luminosity comes from the reactions listed above, then solar luminosity might have been higher by ~1-2% during the critical evolutionary period when the Standard Solar Model predicts frozen oceans and a "faint early Sun" [121]. Circular polarized light from the neutron star may have separated d- and l-amino acids before the appearance of life [101].

Life and subatomic particles evolved together on opposite sides of the photosphere, the brightly glowing layer that is commonly referred to as the Sun or the surface of the Sun. Beneath the photosphere, neutrons from the Sun's compact energetic core became atoms of hydrogen and helium with a $\sim 10^{15}$ fold increase in volume. Above the photosphere, life evolved on an iron-rich planet that orbits in the heliosphere - the outer layer of the Sun. The evolution of life and its survival by adaptation were enhanced by the constantly changing positions of planets around the Sun. This induced changes in the Sun's motion around the centre-of-mass (barycentre) of the solar system, solar cycles and natural changes in Earth's climate [122,123].

## Acknowledgements

This paper is dedicated to my son, Wesley Kyle Manuel, who was born about a month and a half after attending the 1976 AGU meeting [17] in his mother's womb. Words cannot properly express my gratitude to all those who contributed so much to the success of this project. I am most indebted for the loving support of my wife, Caroline Annette Howk Manuel, to other family members, and to literally hundreds of scientists, colleagues, mentors, friends and former students. The Foundation for Chemical Research, Inc., FCR, granted permission to reproduce figures from earlier reports.

## References

1. K. Michaelian, "Thermodynamic origin of life," Earth System Dynamics Discussion, vol. 1, March 2010, pp. 1-39. http://www.earth-syst-dynam-discuss.net/1/1/2010/esdd-1-1-2010.pdf
2. W. A. Fowler, "Forward," In: David N. Schramm, Ed., Cauldrons in the Cosmos: Nuclear Astrophysics by Claus E. Rolf and William S. Rodney, University of Chicago Press, Chicago, IL, USA, 1988, pp. xi-xii. http://www.omatumr.com/Fowler1988/CaldronsCosmos.pdf

3. R. S. Lewis, B. Srinivasan, and E. Anders, “Host phase of a strange xenon component in Allende," Science, vol. 190, 1975, pp. 1251-1262.
4. E. Anders, H. Higuchi, J. Gros, H. Takahashi and J.W. Morgan, "Extinct super-heavy element in the Allende meteorite," Science, vol. 190, 1975, pp. 1262-1271.
5. B2FH: E. M. Burbidge, G. R. Burbidge, W. A. Fowler and F. Hoyle, “Synthesis of the elements in stars,” Reviews of Modern Physics, vol. 29, 1957, pp. 547-650.
6. O. K. Manuel and D. D. Sabu, “Elemental and isotopic inhomogeneities in noble gases: The case for local synthesis of the chemical elements,” Transactions, Missouri Academy of Science, vol. 9, 1975, pp. 104-122.
7. J. H. Reynolds, “Determination of the age of the elements,” Physical Review Letters, vol. 4, January 1960, pp. 8-10. http://prl.aps.org/abstract/PRL/v4/i1/p8_1
8. J. H. Reynolds, “Isotopic composition of primordial xenon,” Physical Review Letters, vol. 4, April 1960, pp. 351-354. http://prl.aps.org/abstract/PRL/v4/i7/p351_1
9. P. K. Kuroda, “Nuclear fission in the early history of the Earth,” Nature, vol. 187, 1960, pp. 36-38. http://www.nature.com/nature/journal/v187/n4731/abs/187036a0.html
10. M. W. Rowe and P. K. Kuroda, “Fissiogenic xenon from the Pasamonte meteorite,” Journal of Geophysical Research, vol. 79, 1965, pp. 709-714. http://www.agu.org/journals/ABS/1965/JZ070i003p00709.shtml
11. E. C. Alexander, Jr. and O. K. Manuel, "Isotopic anomalies of krypton and xenon in Canyon Diablo graphite," Earth and Planetary Science Letters, vol. 2, May 1967, pp. 220- 224. http://tinyurl.com/4rf4h6n
12. M. S. Boulos and O. K. Manuel, “The xenon record of extinct radioactivities in the earth,” Science, vol. 174, 1971, pp. 1334-1336.
13. C. M. Gray and W. Compston, "Excess $^{26}$Mg in the Allende meteorite," Nature, vol. 251, 1974, pp. 495-497.
14. D. D. Clayton, “Extinct radioactivities: Trapped residuals of presolar grains,” Astrophysics Journal, vol. 199, 1975, pp. 765-769.
15. O. K. Manuel, E. W. Hennecke and D. D. Sabu, “Xenon in carbonaceous chondrites,” Nature, vol. 240, 1972, pp. 99-101.
16. R. N. Clayton, L. Grossman and T. K. Mayeda, “A component of primitive nuclear composition in carbonaceous meteorites,” Science, vol. 182, 1973, pp. 485-488.
17. D. D. Sabu and O. K. Manuel, “The xenon record of element synthesis,” Transactions, American Geophysical Union, vol. 57, April 14, 1976, p. 278.
18. O. K. Manuel and D. D. Sabu, “The noble gas record of nucleosynthesis”, paper presented at the 172nd National Meeting of the American Chemical Society, Symposium on the Origin of the Elements, abstract 52, September 2, 1976, San Francisco, CA, USA.
19. O. K. Manuel, Reply to: "A superheavy element in meteorites?" by Dr. Edward Anders, In: W. O. Milligan, Ed., Proceedings of the Robert A. Welch Foundation Conferences on Chemical Research XXI. COSMOCHEMISTRY, The Robert A. Welch Foundation, Houston, TX 77002 USA, November 7, 1977, pp. 263-272.
20. Edward Anders, "Evidence for an extinct super-heavy element in meteorites", paper presented at the 172nd National Meeting of the American Chemical Society, Symposium on the Origin of the Elements, abstract 52, September 2, 1976, San Francisco, CA, USA.

21. Edward Anders, "A super-heavy element in meteorites?" In: W. O. Milligan, Ed., Proceedings of the Robert A. Welch Foundation Conferences on Chemical Research XXI. COSMOCHEMISTRY, The Robert A. Welch Foundation, Houston, TX 77002 USA, November 7, 1977, pp. 245-263.
22. O. K. Manuel and D. D. Sabu, "Strange xenon, extinct super-heavy elements and the solar neutrino puzzle," Science, vol. 195, January 14, 1977, pp. 208-209.
23. R. S. Lewis, B. Srinivasan and E. Anders, "Reply O. K. Manuel and D. D. Sabu," Science, vol. 195, January 14, 1977, pp. 209-210.
24. H. Takahashi, H. Higuchi, J. Gros, J.W. Morgan, and E. Anders, "Allende meteorite: Isotopically anomalous xenon is accompanied by normal osmium," Proceedings of the National Academy of Science USA, vol. 73, 1976, pp. 4253-4256.
25. A. G. W. Cameron and J. W. Truran, "The supernova trigger for formation of the solar system," Icarus, vol. 30, 1977, pp. 447-461.
26. R. N. Clayton, N. Onuma and T. K. Mayeda, "A classification of meteorites based on oxygen isotopes," Earth and Planetary Science Letters, vol. 30, April 1976, pp. 10-18.
27. R. N. Clayton and T. K. Mayeda, "Correlated oxygen and magnesium isotope anomalies in Allende inclusions, I: Oxygen", Geophysical Research Letters, vol. **4**, 1977, pp. 295-298.
28. G. J. Wasserburg, T. Lee and D. A. Papanastassiou, "Correlated O and Mg isotope anomalies in Allende inclusions, II: Magnesium", Geophysical Research Letters, vol. 4, 1977, pp. 299-302.
29. D. D. Clayton, E. Druck, and S. E. Woosley, "Isotopic anomalies and proton irradiation in the early solar system," Astrophysics Journal, vol. 214, 1977, pp. 300-315.
30. M. T. McCulloch and G. J. Wasserburg, "Barium and neodymium isotopic anomalies in the Allende meteorite," Astrophysics Journal, vol. 220, February 15, 1978, pp. L15-L19.
31. T. Lee, D. A. Papanastassiou and G. J. Wasserburg, "Calcium isotopic anomalies in the Allende meteorite," Astrophysics Journal, vol. 220, 1978, pp. L21-L25.
32. M. T. McCulloch and G. J. Wasserburg, "More anomalies from the Allende meteorite: Samarium," Geophysical Research Letters, vol. 5, 1978, pp. 599-602.
33. G. W. Lugmair, K. Marti and N. B. Scheinin, "Incomplete mixing of products from r-, p-, and s-process nucleosynthesis: Sm-Nd systematics in Allende inclusions EK 1-04-1," Lunar and Planetary Science Abstracts, vol. IX, 1978, pp. 672-674.
34. D. A. Papanastassiou, J. C. Huneke, T. M. Esat and G. J. Wasserburg, "Pandora's box of nuclides," Lunar and Planetary Science Abstracts, vol. IX, 1978, pp. 859-861.
35. R. V. Ballad, L. L. Oliver, R. G. Downing and O. K. Manuel, "Nucleogenetic heterogeneities in chemical and isotopic abundances", Meteoritics, vol. 13, 1978, pp. 387-391.
36. B. Srinivasan B. and E. Anders, "Noble gases in the Murchison meteorite: Possible relics of s-process nucleosynthesis", Science, vol. 201, 1978, pp. 51-56.
37. U. Frick and S. Chang, "Elimination of chromite and novel sulfides as important carriers of noble gases in carbonaceous meteorites", Meteoritics, vol. 13, 1978, pp. 465-469.
38. W. R. Kelly and G. J. Wasserburg, " Evidence for the existence of $^{107}$Pd in the early solar system," Geophysical Research Letters, vol. 5, 1978, pp. 1079-1082.

39. R. S. Lewis, L. Alaerts, J -I. Matsuda and E. Anders, "Stellar condensates in meteorites: Isotopic evidence from noble gases," Astrophysics Journal, vol. 234, 1979, pp. L165-Ll68
40. L. L. Oliver, R. V. Ballad, J. F. Richardson and O. K. Manuel, "Correlated anomalies of tellurium and xenon in Allende", Meteoritics, vol. 14, 1979, pp. 503-506.
41. P. K. Kuroda, "Isotopic anomalies in the early solar system," Geochemical Journal, vol. 13, 1979, pp. 83-90. http://www.terrapub.co.jp/journals/GJ/pdf/1302/13020083.PDF
42. R. V. Ballad, L. L. Oliver, R. G. Downing and O. K. Manuel, "Isotopes of tellurium, xenon, and krypton in Allende meteorite retain record of nucleosynthesis", Nature**,** vol. 277, February 1979, pp. 615-620. http://www.nature.com/nature/journal/v277/n5698/abs/277615a0.html
43. S. Niemeyer, "I-Xe dating of silicate and troilite from IAB iron meteorites", Geochimica et Cosmochimica Acta, vol. 43, 1979, pp. 843-860.
44. O. K. Manuel, "The enigma of helium and anomalous xenon", Icarus, vol. 41, February 1980, pp. 312-315. http://tinyurl.com/4qpcz92
45. F. Begemann, "Isotopic anomalies in meteorites," Reports on Progress in Physics, vol. 43, Nov. 1980, pp. 1309-1356. http://iopscience.iop.org/0034-4885/43/11/002
46. D. D. Sabu and O. K. Manuel, "Noble gas anomalies and synthesis of the chemical elements", Meteoritics, vol**.** 15, June 1980, pp. 117-138. http://adsabs.harvard.edu/full/1980Metic..15..117S
47. K. J. R. Rosman, J. R. DeLaeter and M. P. Gorton, "Cadmium isotope fractionation in fractions of two H3 chondrites," Earth and Planetary Science Letters, vol. 48, June 1980, pp. 166-170. http://tinyurl.com/4pctn4v
48. A. K. Lavrukhina, "On the nature of isotopic anomalies in meteorites," Nukleonika, vol. 25, 1980, pp. 1495-1515.
49. Typhoon Lee and M. L. Zeff, "A Mg survey for more "FUN" isotopic anomalies," Lunar and Planetary Science, vol. XII, March 1981, pp. 604-606.
50. L. L. Oliver, R. V. Ballad, J. F. Richardson and O. K. Manuel, "Isotopically anomalous tellurium in Allende: Another relic of local element synthesis," Journal of Inorganic and Nuclear Chemistry, vol. 43, 1981, pp. 2207-2216. http://tinyurl.com/4n5wdlu
51. O. K. Manuel, "Heterogeneity of isotopic and elemental compositions in meteorites: Evidence of local synthesis of the elements," Geokhimiya, number 12, 1981, pp. 1776-1801.
52. O. K. Manuel and D. D. Sabu, "The noble gas record of the terrestrial planets", Geochemical Journal**,** vol. 15, 1981, pp. 245-267. http://www.terrapub.co.jp/journals/GJ/pdf/1505/15050245.PDF
53. G. J. Mathews amd W. A. Fowler, "Systematics of r-process enrichment factors for barium, neodynium, and samarium isotopic anomalies in the Allende meteorite," Astrophysical Journal, vol. 251, December 1981 pp. L45-L48. http://adsabs.harvard.edu/full/1981ApJ...251L..45M
54. A. K. Lavrukhina and R. I. Kuznetsova, "Irradiation effects at the supernova stage in isotopic anomalies," Lunar and Planetary Science Abstracts, vol. XIII, March 1982, pp. 425-426.

55. P. K. Kuroda, "The origin of chemical elements and the Oklo phenomenon," Springer-Verlag, New York, NY, July 1982. http://www.amazon.com/Origin-Chemical-Elements-Oklo-Phenomenon/dp/3540116796
56. G. Hwaung and O. K. Manuel, "Terrestrial-type xenon in meteoritic troilite," Nature, vol. 299, October 1982, pp. 807-810. http://www.nature.com/nature/journal/v299/n5886/abs/299807a0.html
57. K. K. Turekian and S. P. Clarke, "Inhomogeneous accumulation of the earth from the primitive solar nebula," Earth and Planetary Science Letters, vol. 6, 1969, pp. 346-348.
58. A. P. Vinogradov, "Formation of the metal cores of planets," Geokhimiya, vol. 10, 1975, pp. 1427-1431.
59. W. A. Fowler, J. L. Greenstein and F. Hoyle, "Nucleosynthesis during the early history of the solar system," Geophysical Journal of the Royal Astronomical Society, vol. 6, 1962, pp. 148-220.
60. P. Signer and H. Suess,"Rare gases in the sun, in the atmosphere and in meteorites," In J. Geiss and E. D. Goldberg, Ed., Earth Science and Meteorites, North Holland, Amsterdam, 1963, pp. 241-272.
61. R. O. Pepin and P. Signer, "Primordial rare gases in meteorites," Science, vol. 149, 1965, pp. 253-265.
62. O. K. Manuel, "Noble gases in the Fayetteville meteorite", Geochimica et Cosmochimica Acta, vol. 31, 1967, pp. 2413-2431.
63. K. Marti, "Solar-type xenon: A new isotopic composition of xenon in the Pesyanoe meteorite," Science, vol. 166, 1969, pp. 1263-1265.
64. P. K. Kuroda and O. K. Manuel, "Mass fractionation and isotope anomalies in neon and xenon", Nature, vol. 227, 1970, pp. 1113-1116.
65. O. K. Manuel, R. J. Wright, D. K. Miller and P. K. Kuroda, "Heavy noble gases in Leoville: The case for mass fractionated xenon in carbonaceous chondrites", Journal of Geophysical Research, vol. 75, 1970, pp. 5693-5701.
66. E. W. Hennecke and O. K. Manuel, "Mass fractionation and the isotopic anomalies of xenon and krypton in ordinary chondrites", Zeitschrift für Naturforschung, vol. 20a, 1971, pp. 1980-1986.
67. B. Srinivasan and O. K. Manuel, "On the isotopic composition of trapped helium and neon in carbonaceous chondrites", Earth and Planetary Science Letters, vol. 12, 1971, pp. 282-286.
68. B. Srinivasan, E. W. Hennecke, D. E. Sinclair and O. K. Manuel, "A comparison of noble gases released from lunar fines (#15601.64) with noble gases in meteorites and in the Earth," Proceedings of the Third Lunar Science Conference (Supplement 3, Geochimica et Cosmochimica Acta), vol. 2, Houston, TX, March 1972, pp. 1927-1945.
69. D. D. Sabu and O. K. Manuel, "The neon alphabet game," Lunar and Planetary Science Abstract, vol. XI, 1980, pp. 971-973. http://adsabs.harvard.edu/full/1980LPI....11..971S
70. D. D. Sabu and O. K. Manuel, "The neon alphabet game", Proceedings Eleventh Lunar and Planetary Science Conference, vol. 2, March 1980, pp. 879-899. http://adsabs.harvard.edu/full/1980LPSC...11..879S

71. R. S. Lewis, E. Anders, T. Shimamura, and G. W. Lugnair, "Search for isotopic anomalies correlated with CCF xenon: I. Barium", Lunar and Planetary Science Abstract #1221, vol. XIV, March 1983, pp. 436-437. http://www.lpi.usra.edu/meetings/lpsc1983/pdf/1221.pdf
72. I. P. Wright, S. J. Norris, C. T. Pillinger, R. S. Lewis and E. Anders, "Light nitrogen, its composition, association and location in the Allende and Murchison meteorites," Lunar and Planetary Science Abstract #1436, vol. XIV, March 1983, pp. 861-862. http://www.lpi.usra.edu/meetings/lpsc1983/pdf/1436.pdf
73. O. K. Manuel and G. Hwaung, "Information of astrophysical interest in the isotopes of solar-wind implanted elements," Lunar and Planetary Science Abstract #1232, vol. XIV, March 1983, pp. 458-459. http://db.tt/BNB6bnp http://www.lpi.usra.edu/meetings/lpsc1983/pdf/1232.pdf
74. P. K. Swart, "The demise of established dogmas on the formation of the Solar System", Nature, vol. 303, 26 May 1983, pp. 286-286. http://tallbloke.files.wordpress.com/2010/01/swart-1983.pdf
75. O. K. Manuel and G. Hwaung, "Solar abundances of the elements," Meteoritics, vol.18, September 1983, pp. 209-222. http://tinyurl.com/224kz4
76. P. Eberhardt, J. Geiss, H. Graf, N. Grögler, M. D. Mendia, M. Mörgeli, H. Schwaller, A. Stettler, U. Krähenbühl and H. R. von Guten, "Trapped solar wind noble gases in Apollo 12 lunar fines 12001 and Apollo 11 breccia 10046," Proceedings of the Third Lunar Science Conference (Supplement 3, Geochimica et Cosmochimica Acta), vol. 2, Houston, TX, March 1972, pp. 1821-1856.
77. A. O. Nier, "A re-determination of the relative abundances of the isotopes of neon, krypton, rubidium, xenon and mercury," Physical Review, vol. 79, 1950, pp. 450-454.
78. U. Frick, "Anomalous krypton in the Allende meteorite," Proceedings of the Eighth Lunar Science Conference (Lunar Science Institute, Houston, TX 77058), vol. 1, Houston, TX, March 1977, pp. 273-292.
79. J. H. Reynolds, U. Frick, J. M. Neil, and D. L. Phinney, "Rare-gas-rich separates from carbonaceous chondrites", Geochimica et Cosmochimica Acta, vol. 42, 1978, pp. 1775-1797.
80. J. E. Ross and L. H. Aller, "The chemical composition of the Sun," Science, vol. 191, 1976, pp. 1223-1229.
81. W. D. Harkins, "The evolution of the elements and the stability of complex atoms", Journal of the American Chemical Society, vol. 39, 1917, pp. 856-879.
82. R. S. Lewis, E. Anders, T. Shimamura, and G. W. Lugmair, "Barium isotopes in Allende meteorite: Evidence against an extinct superheavy element", Science, vol. 222, December 1983, pp. 1013-1015. http://www.sciencemag.org/content/222/4627/1013.abstract
83. Peter Toth, "Is the Sun a pulsar?" Nature, vol. 270, November 1977, pp. 159-160. http://www.nature.com/nature/journal/v270/n5633/abs/270159a0.html
84. U. Ott and F. Begemann, "Discovery of s-process barium in the Murchison meteorite", The Astrophysical Journal, vol. 353, 1990, pp. L57 - L60.
85. F. Begemann, "Isotopic abundance anomalies and the early solar system," In N. Prantos, E. Vangioni-Flam and M Cassé, Eds., Origin and Evolution of the Elements, Cambridge University Press, Cambridge, UK, 1993, pp. 518-527.
86. J. T. Lee, Bin Li and O. K. Manuel, "Terrestrial-type xenon in sulfides of the Allende meteorite", Geochemical Journal, vol. 30, 1996, pp. 17-30.

87. J. T. Lee, Bin Li and O. K. Manuel, "On the signature of local element synthesis", Comments on Astrophysics, vol. 18, 1997, pp. 335-345. [Listed in the Smithsonian/NASA Astrophysics Data System: Communications in Astrophysics, vol. 18, 1996, pp. 335-345] http://adsabs.harvard.edu/abs/1996ComAp..18..335L
88. G. K. Nicolussi, A. M. Davis, M. J. Pellin, R. S. Lewis, R. N. Clayton and S. Amari, "s-Process zirconium in presolar silicon carbide grains", Science, vol. 277, 1997, pp. 1281-1283.
89. G. K. Nicolussi, M. J. Pellin, R. S. Lewis, A. M. Davis, S. Amari and R. N. Clayton, "Molybdenum isotopic composition of individual presolar silicon carbide grains from the Murchison meteorite, Geochimica et Cosmochimica Acta, vol. 62, 1998, 1093-1104.
90. G. K. Nicolussi, A. M. Davis, M. J. Pellin, R. S. Lewis, R. N. Clayton and S. Amari, "Strontium isotopic composition in individual circumstellar silicon carbide grains: A record of s-process nucleosynthesis, Physical Review Letters, vol 81, 1998, pp. 3583-3586.
91. S. Richter, U. Ott, and F. Begemann, "Tellurium in presolar diamonds as an indicator for rapid separation of supernova ejecta", Nature, vol. 391, 1998, pp. 261-263.
92. O. K. Manuel, J. T. Lee, D. E. Ragland, J. M. D. MacElroy, Bin Li and Wilbur Brown, "Origin of the solar system and its elements", Journal of Radioanalytical and Nuclear Chemistry, vol. 238, 1998, pp. 213-225.
93. R. Maas, R.D. Loss, K.J.R. Rosman, J.R. DeLaeter, U. Ott, R.S. Lewis, G. H. Huss, E. Anders, and G. W. Lugmair, "Isotope anomalies in tellurium in interstellar diamonds", In O. Manuel, Ed., Origin of Elements in the Solar System: Implications of Post 1957 Observations, Kluwer Academic/Plenum Publishers, New York, 2000, pp. 361-367.
94. Ulrich Ott, "Isotope abundance anomalies in meteorites: Clues to yields of individual nucleosynthesis processes", In O. Manuel, Ed., Origin of Elements in the Solar System: Implications of Post 1957 Observations, Kluwer Academic/Plenum Publishers, New York, 2000, pp. 361-367.
95. C. A. Rouse, "Evidence for a small, high-Z, iron-like solar core," Astronomy and Astrophysics, vol. 149, August 1985, pp. 65-72. http://articles.adsabs.harvard.edu/full/1985A%26A...149...65R
96. O. Manuel, "Isotope ratios in Jupiter confirm intra-solar diffusion", *Meteoritics and Planetary Science* **33**, A97, abstract 5011 (1998).
97. D. V. Reames, "Abundances of trans-iron elements in solar energetic particle events," The Astrophysical Journal, vol. 540, September 2000, pp. L111 - L114. http://epact2.gsfc.nasa.gov/don/00HiZ.pdf
98. O. Manuel, C. Bolon, A. Katragada and M. Insall, "Attraction and repulsion of nucleons: Sources of stellar energy," Journal of Fusion Energy, vol. 19, March 2000, pp. 93-98.
99. O. Manuel, C. Bolon, M. Zhong, and P. Jangam, "The sun's origin, composition and source of energy," Lunar and Planetary Science Abstracts, vol. XXXII, March 2001, abstract 1041, http://arxiv.org/pdf/astro-ph/0411255v1
100. T. Kirsten, "Solar neutrino experiments: Results and implications," Reviews of Modern Physics, vol. 71, 1999, pp. 1213-1232.
101. J. R. Cronin and S. Pizzarello, "Enantiomeric excesses in meteoritic amino acids," Science, vol. 274, 14 February 1997, pp. 951-955.

http://www.sciencemag.org/content/275/5302/951.abstract

102. O. Hahn and F. Strassmann, "Über den Nachweis und das Verhalten der bei der Bestrahlung des Urans mittels Neutronen entstehenden Erdalkalimetalle (On the detection and characteristics of the alkaline earth metals formed by irradiation of uranium with neutrons)," Naturwissenschaften, vol. 27, no. 1, 1939, pp. 11–15.
103. J. K. Tuli, *Nuclear Wallet Cards,* Sixth Edition, Upton, NY: National Nuclear Data Center, Brookhaven National Lab, 2000, 74 pp.
104. H. Heiselberg, "Neutron star masses, radii and equation of state", in *Proceedings of the Conference on Compact Stars in the QCD Phase Diagram,* eConf C010815, edited by R. Ouyed and F. Sannion, Copenhagen, Denmark, Nordic Institute for Theoretical Physics (2002) pp. 3-16, http://www.arxiv.org/abs/astro-ph/?0201465
105. N. Rea, P. Esposito, R. Turolla, G. L. Israel, S. Zane, L. Stella, S. Mereghetti, A. Tiengo, D. Götz, E. Göğüş and C. Kouveliotou, "A low-magnetic-field soft gamma repeater," Science, vol. 330, November 2010, pp. 944-946. http://www.sciencemag.org/content/330/6006/944.abstract
106. A. A. Abdo *et al.* (~170 co-authors), "Gamma-ray flares from the Crab Nebula," Science, (ScienceXPress) Published online 6 January 2011, [DOI:10.1126/science.1199705] http://www.sciencexpress.org http://dl.dropbox.com/u/10640850/Science-2011-Abdo-science.1199705.pdf
107. University College London, "Mysterious pulsar with hidden powers discovered," reported in *ScienceDaily*, 15 October 2010, Retrieved on January 14, 2011 from http://www.sciencedaily.com/releases/2010/10/101014144231.htm
108. O. Manuel, E. Miller and A. Katragada, "Neutron repulsion confirmed as energy source," Journal of Fusion Energy, vol. 20, December 2001, pp. 197-201. http://www.springerlink.com/content/x1n87370x6685079/
109. O. Manuel, C. Bolon and M. Zhong, "Nuclear systematics Part III: The source of solar luminosity", Journal of Radioanalytical and Nuclear Chemistry, vol. 252, 2002, pp. 3-7.
110. D. Lunney, J. M. Pearson and C. Thibault, "Recent trends in the determination of nuclear masses," Reviews of Modern Physics, vol. 75, 2003, pp. 1021-1082.
111. W. Baade and F. Zwicky, "Cosmic rays from super-novae", Proceedings of the National Academy of Sciences, vol. 20, 1934, pp. 259-263.
112. J. R. Oppenheimer and G. M. Volkoff, "On massive neutron cores", Physical Review, vol. 15, 1939, pp. 374-381.
113. Q. R. Ahmad *et al.* (~180 co-authors), "Measurement of charged current interactions produced by $^8$B solar neutrinos at the Sudbury Neutrino Observatory," Physical Review Letters, vol. 87, July 2001 (6 pages) 71301 http://arxiv.org/pdf/nucl-ex/0106015v2 http://prl.aps.org/abstract/PRL/v87/i7/e071301
114. Q. R. Ahmad *et al.* (~180 co-authors), "Direct evidence for neutrino flavor transformation from neutral-current interactions in the Sudbury Neutrino Observatory," Physical Review Letters, vol. 89, no. 1, 2002 (6 pages) 011301: http://arxiv.org/pdf/nucl-ex/0204008v2 http://prl.aps.org/abstract/PRL/v89/i1/e011301
115. A. A. Aguilar-Arevalo *et al.* (~153 co-authors), "Event excess in the MiniBooNE search for $\bar{\nu}_\mu \rightarrow \nu_e$ oscillations," Physical Review Letters, vol. 105, 2010 (5 pages) 0181801: http://prl.aps.org/abstract/PRL/v105/i18/e181801

116. N. C. Moore, “Physics experiment suggests existence of new particle,” University of Michigan News Service, 1 November 2010. http://ns.umich.edu/htdocs/releases/story.php?id=8076
117. A. Frank, Bipolar outflows in stellar astrophysics,” In O. Manuel, Ed., Origin of Elements in the Solar System: Implications of Post 1957 Observations, Kluwer Academic/Plenum Publishers, New York, 2000, pp. 241-249.
118. O. K. Manuel, B. W. Ninham and S. E. Friberg, "Superfluidity in the solar interior: Implications for solar eruptions and climate," Journal of Fusion Energy, vol. 21, 2003, pp. 193-198. http://www.springerlink.com/content/r2352635vv166363/
119. K. Birkeland, “The Norwegian Aurora Polaris Expedition 1902-1903,” H. Aschehoug & Co, 1908. http://www.archive.org/details/norwegianaurorap01chririch http://www.archive.org/stream/norwegianaurorap01chririch - page/n5/mode/2up
120. M. Mozina, H. Ratcliff and O. Manuel, "Observational confirmation of the Sun's CNO cycle," Journal of Fusion Energy, vol. 25, Oct 2006, pp. 141-144. http://arxiv.org/pdf/astro-ph/0512633v2
121. M. T. Rosing, D. K. Bird, N. H. Sleep and C. J. Bjerrum, "No climate paradox under the faint early Sun", Nature, vol. 464, April 2010, pp. 744-747.
122. P. D. Jose, “Sun’s motion and sunspots”, Astronomical Journal, vol. 70, 1965, pp. 193-200.
123. R. Mackey, “Rhodes Fairbridge and the idea that the solar system regulates the Earth’s climate,” (Proceedings of the Ninth International Coastal Symposium, Gold Coast, Australia), Journal of Coastal Research, SI 50, 2007, pp. 955-968. http://www.griffith.edu.au/conference/ics2007/pdf/ICS176.pdf